\documentclass[
journal=ancac3, 
manuscript=article]{achemso}

\usepackage{subcaption}
\usepackage{rotating}
\usepackage{comment}
\usepackage[version=4]{mhchem}
\usepackage{multirow}
\usepackage{siunitx,booktabs}

\author{Shivam Parashar}
\author{Prateek Kumar Jha}
\email{pkjchfch@iitr.ac.in}

\affiliation[IIT R]
{Department of Chemical Engineering, Indian Institute of Technology Roorkee}

\title[\texttt{achemso} demonstration]
{\ Trends in pK$_a$ Values for polyprotic carboxylic acids}

\begin{document}

\setstretch{1.3}
\begin{abstract}
In this work we have reported the pK$_a$ values for straight chain (withouot branches) oligomers of Acrylic acid using CBS-QB3 gas phase energies with solvation free energies from CPCM/HF 6-31G(d) and SMD/B3LYP (2df,p). We have found that the negative log of the first acid dissociation constant decreases with the chain length. Within a chain, the microscopic pK$_a$ for the COOH groups present at the end of the chain is maximum and minimum for centrally located COOH groups.
\end{abstract}

\section{Introduction}

A lot of research has been conducted to model and predict the pK$_a$ values for small inorganic acids using quantum mechanics~\cite{doi:10.1021/ja010534f}. Effect of thermodynamic cycle~\cite{silva}, model chemistries~\cite{silva}, implicit and explicit solvation~\cite{silva} and solute cavity~\cite{lee2011} on the accuracy of the pK$_a$ values has also been investigated. There has been fewer studies~\cite{anil2015,lee2011,metha} for investigation of acidity constants for polymeric acids because of presence of multiple acid groups. These acids have a range of applications in specific bioseparation and catalysis~\cite{metha2}. In this study we have considered the oligomers of acrylic acid, in which a single kind of functional group (COOH) is present, and the microscopic pKa value of each site is dependent on its location relative to the molecule. In case of a protein, the same -COOH group could have a pKa value of 2 and 9 units~\cite{protein} indicating the significance of the electrostatic environment.

The high-level ab initio CBS-QB3 and CBS-APNO methods (using HF 6-31 G(d) and HF 6-31+ G(d)) with CPCM generated smallest inaccuracy of about 0.5 log unit when commonly used, and the simplest thermodynamic cycle (without involving water molecule) was employed~\cite{Liptak2001}.

Qian et al.~\cite{metha} have reported the pK$_a$ values for different conformers for dimer and trimer of methaacrylic acid. They found the pK$_a$ values of the oligomers to be greater than the monomer. But they had not considered the anionic structures in which intramolecular H-bonding could stablize the structure thereby facilitating easy acid dissociation.

\section{Theory}

The Continuum solvent models~\cite{tomasi2005} have been parametrized to give accurate solvation energies at low levels of theory. The accuracy of continuum solvent calculations of aqueous pKas lies in the range of 2-3 units~\cite{Ho2009}. For accurately calculating the free energy in the solvent, the solvation energy should be combined with the high-level ab initio methods for the gas-phase which is accomplised by using an appropriate thermodynamic cycle. We have used a Direct thermodynamic cycle to compute $\Delta G_{aq}$. The direct cycle has provided errors of approximately 0.5 pK$_a$ units or lower for carboxylic acids ~\cite{Liptak2001}. $\Delta G_{aq}$ is given by -

\begin{equation}
\Delta G_{aq} = \Delta G_{gas} + \Delta \Delta G_{sol}
\end{equation}

Where  $\Delta G_{gas}$ is given by -

\begin{equation}
\Delta G_{gas} = G_{gas}(H^+) + G_{gas}(A^-) - G_{gas}(HA)
\label{G-gas}
\end{equation}

and $\Delta \Delta G_{sol}$ can be written as - 

\begin{equation}
\Delta \Delta G_{sol} = \Delta G_{sol}(H^+) + \Delta G_{sol}(A^-) - \Delta G_{sol}(HA)
\label{G-sol}
\end{equation}

All the terms in RHS of equation~\ref{G-gas} and ~\ref{G-sol} were computed except $G_{gas}(H^+)$ and $\Delta G_{sol}(H^+)$ as free energy cannot be calculated for a proton since it does not contains any electrons. Therefore, those values were taken from experiments.

After assuming that the thermal contributions are similar in both the gas and solution phase, each term of $\Delta G_{sol}$ in equation~\ref{G-sol} was calculated as~\cite{ho2010} -
\begin{equation}
\Delta G_{sol} = (E_{soln}+ G_{nes}) - E_{gas}
\end{equation}

where $E_{soln}$ and $E_{gas}$ are the electronic energy of molecule in presence and absence of the solvent field respectively. $G_{nes}$ is the sum of all non-electrostatic contributions which includes cavitation and dispersion-repulsion terms. In Gaussian 09, the single point energy obtained using SMD solvent phase calculations contains the non-electrostatic terms in itself. Therefore, solvation energy using SMD model is simply computed as the difference between electronic energy on solution-phase and gas-phase optimized geometries.

For acidic sites present in an oligomer of acrylic acid, the dissociation of the proton from each site cannot be assumed to be stepwise. This case is dissimilar to the classic treatment of a polyprotic acid, for which difference in succesive pKa is large enough to ignore the second or third deprotonation. The equilibriun dissociation constants of individual acidic groups are referred to as microscopic dissociation constants. Their relation to macroscopic dissociation constant, which represents the ability of the whole molecule to lose a proton is given by - 

\begin{equation}
K_1 = k_x + k_y
\end{equation}

where $k_x$ and $k_y$ are the microscopic dissociation constant for acidic sites x and y respectively in a dimer of polyprotic acid, and $K_1$ is the first macroscopic dissociation constant for the dimer.

\section{Simulations}

All simulations were performed on Gaussian 09~\cite{g09} program. Starting geometries for anions were prepared by deleting the appropriate hydrogen atom from their corresponding protonated forms. Several starting geometries were taken and only the most stable conformation energies are reported. It was assumed that the most stable conformer was the one with the least single point energy. Since for each oligomer, there are multiple possible low lying confermers, we optimized   the molecule with different starting structures.

A frequency calculation was also done after the optimization to insure that the no imaginary frequencies are there and the structure is optimized to a true minimum state. In case the high level hessian matrix did not optimized the structue to a true minima, the structure was reoptimized taking the previous unoptimized structure as its initial guess.

The SMD model has been parametrized with a training set of 2821 solvation data including 112 aqueous ionic solvation free energies, 220 solvation free energies for 166 ions in acetonitrile, methanol, and dimethyl sulfoxide, 2346 solvation free energies for 318 neutral solutes in 91 solvents (90 nonaqueous organic solvents and water), and 143 transfer free energies for 93 neutral solutes between water and 15 organic solvents~\cite{SMD}.\\ 

\textbf{Gas-phase energy Calculation}\\

There are a bunch of methods available in gaussian for calculating accurate gas phase energies. Some of them are G1, G2, G3, G4, G2(MP2), G3(MP2) etc. All these methods involve doing several predefined calculations for energy computation. CBS-QB3 is equally accurate and computationally efficient than the above methods.

We have used CBS-QB3~\cite{cbs-qb3} model chemistry to compute $\Delta G_{gas}$ values. It produces mean absolute deviations of 1.43 pka units~\cite{cbs1} for small molecules. CCSD(T) calculations are most accurate and they can give gas-phase free energies as close to the experimental values. But because of their huge computational time, its application is limited to smaller molecules only. CBS-QB3 makes use of optimized geometry from density functional theory and then Single point calculations are performed at various levels such as CCSD(T), MP4SDQ, and MP2~\cite{doi:10.1021/ja010534f}.\\

\textbf{Solvation energy Calculation}
\\

In an earlier study by Liptak and Shields, the authors
showed that calculations using continuum models on the
lowest energy gas-phase conformer and the conformationally
averaged structure gave comparable results~\cite{doi:10.1021/ja010534f} Accordingly, in this study we only consider the solvation free energies on
the lowest energy gas-phase conformer. We have used implicit solvent model in which the solvent is represented  as a polarizable dielectric continuum. Explicit treatment of solvent molecules would be extremely expensive for larger molecule that we are dealing here with. The conductor-polarizable continuum model (CPCM) ~\cite{doi:10.1002/jcc.10189} was used to compute solvation free energies at the HF/6-31 G(d) and SMD was used alongside with B3LYP/6-31 G(2df,p). All geometries of the studied species have been optimized fully in the presence of solvent and their respective level of theories. The implicit model is computationally less expensive than explicit solvent as it approximates the response of the bulk solvent using a homogeneous dieletric continuum. Implicit model does not provides accurate solvation energies for anionic strutures because it does not takes into account individual solute-solvent interaction. However, it performs well for neutral species. In Gaussian 09 the default option is Radii=UA0 (Topological United Atoms model) which treats functional groups as a single sphere. We have optimized the geometry at gas phase and solution phase as well because it increases the accuracy of the pKa values ~\cite{Nina}. Solvation energy was computed as the difference between single point energy in solution phase and in gas phase.

\section{Results}

\begin{table}[h!]
\centering
\hspace*{-1cm}
\begin{tabular}{|c| c |c |c |c |c |c | c|}

 \hline
 \multirow{3}{4em}{Molecule}
 & \multirow{3}{5em}{$\Delta$G$_{gas}$ CBS-QB3 (kcal/mol)} & \multicolumn{4}{c|}{$\Delta$G$_{Sol}$ (kcal/mol) } & \multicolumn{2}{c|}{Relative pK$_a$ } \\
 
 \cline{3-8}
 
 &    & \multicolumn{2}{c|}{HF/6-31G(d)//CPCM} & \multicolumn{2}{|c|}{B3LYP/6-31G(2df,p)//SMD}  & \multirow{2}{3em}{HF} & \multirow{2}{3em}{B3LYP} \\
 
 \cline{3-6}
 &	  & Anion & Acid & Anion & Acid & &
 
 \\
 
 \hline
 
 Propanoic acid & 345.68 & -65.88 & -5.02 & -71.20 & -4.97 & 4.87* & 4.87* \\

& & & & & & & \\
 
 Dimer(Anion-1) & 324.447 & -57.21 & -9.75 & -58.70 & -11.34 & -0.87 & 3.14\\
 Dimer(Anion-3)& 324.452 & -56.55 & -9.75 & -58.03 & -11.34  & -0.39 & 3.63 \\ 
& & & & & & & \\
 Trimer(Anion-1) & 322.41 & -59.63 & -13.99 & -62.25 & -17.99 	 & -1.03 & 3.91\\
 Trimer(Anion-3) & 313.41 & -53.02 & -13.99 & -52.29 & -17.99   & -2.79 & 4.62\\
 Trimer(Anion-5) & 322.91 & -58.73 & -13.99 & -60.40 & -17.99  & 0.00 & 5.64\\
 \hline
 
\end{tabular}
\caption{Summary of the gas-phase and solvation energies for oligomers of Acrylic acid.}
\label{pka}
\end{table}

\begin{figure}[h!]

\begin{subfigure}{.3\textwidth}
  \includegraphics[scale=.20,clip=true, trim = 10mm 10mm 10mm 30mm]{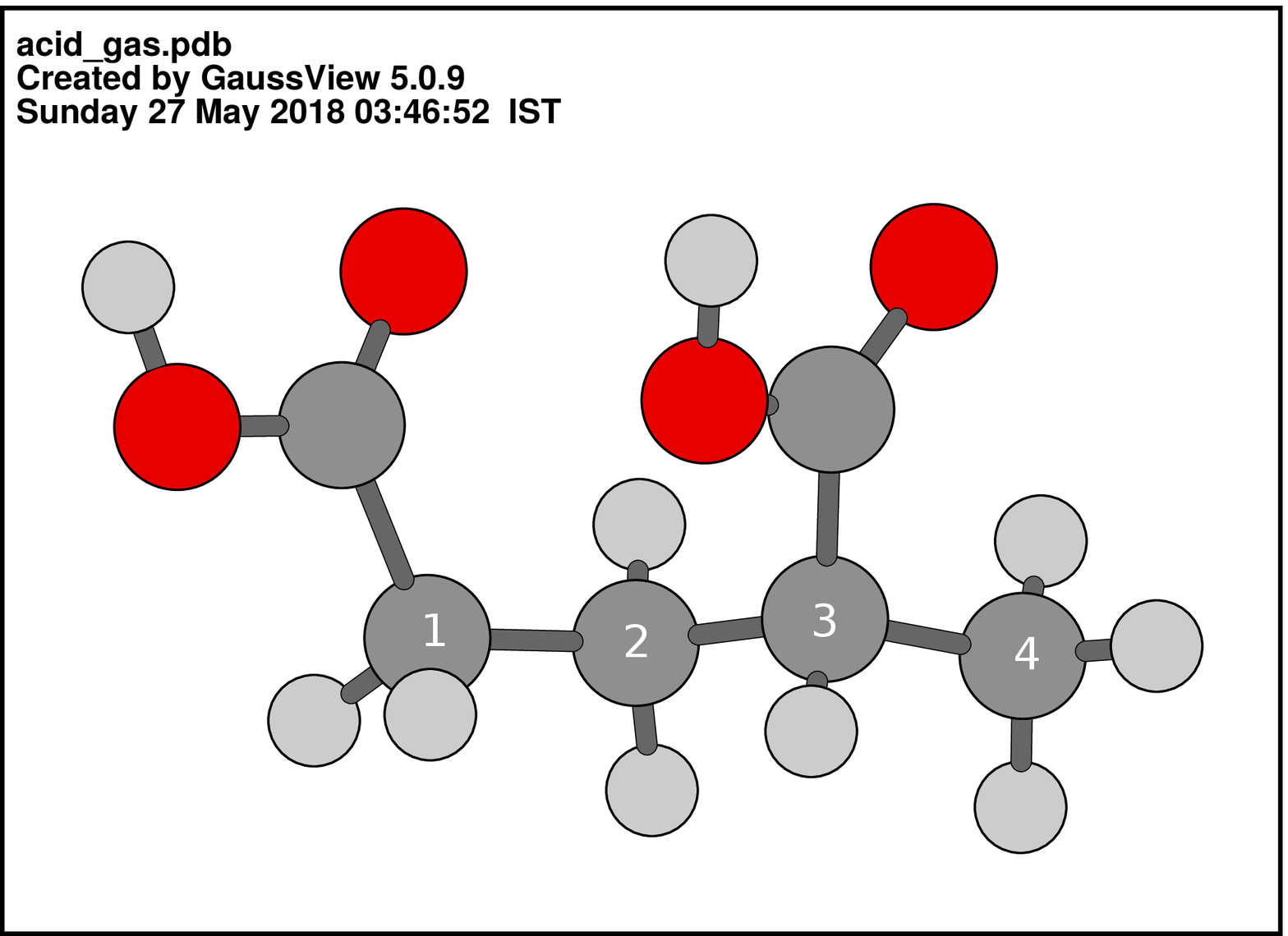} 
   \caption{Acid}\label{dimer_acid}
\end{subfigure}
\begin{subfigure}{.3\textwidth}  
  \includegraphics[scale=.20,clip=true, trim = 10mm 10mm 10mm 30mm]{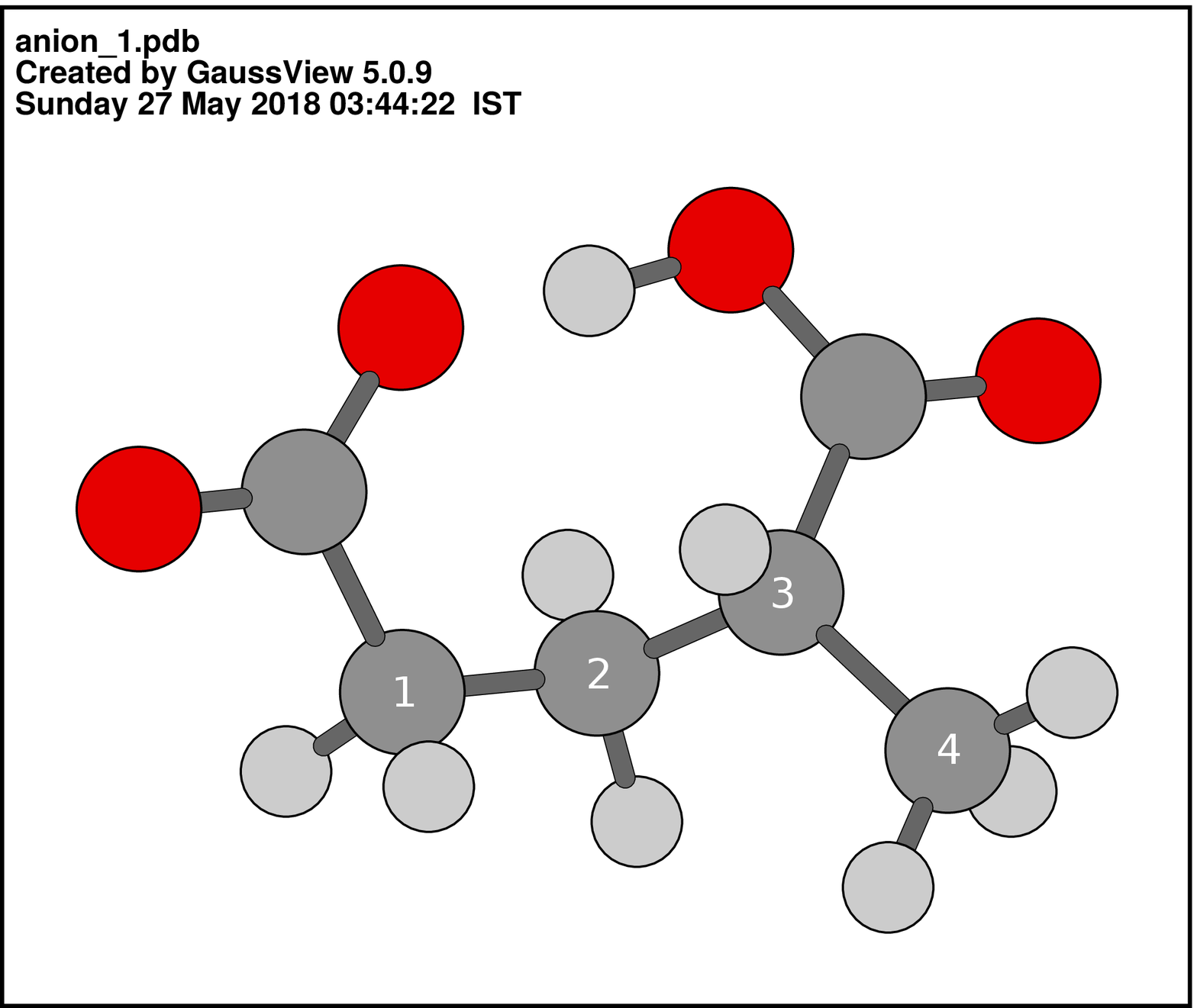}
   \caption{Anion-1}\label{dimer-anion-1}
\end{subfigure}
\begin{subfigure}{.3\textwidth}
\includegraphics[scale=.20,clip=true, trim = 10mm 10mm 10mm 30mm]{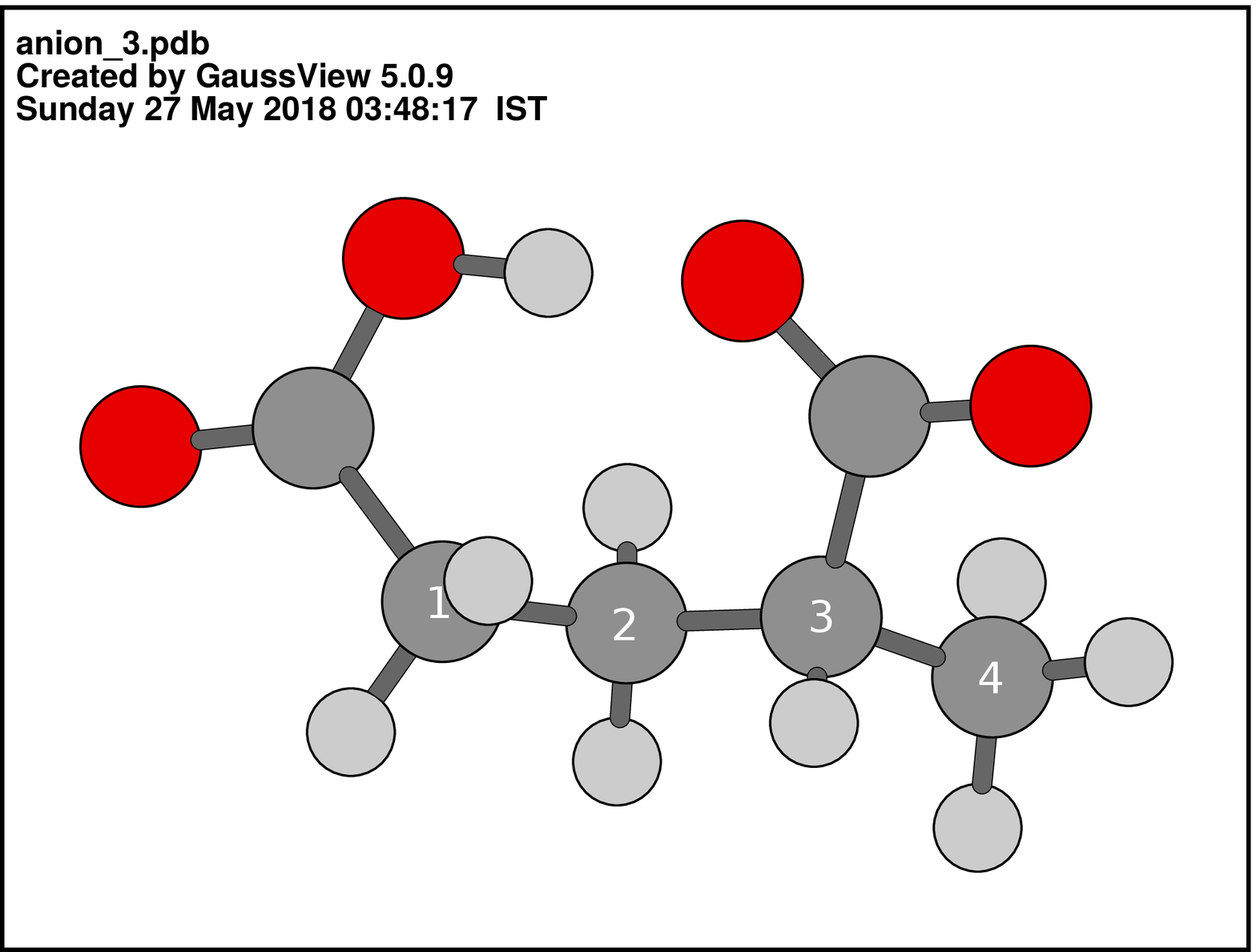}
 \caption{Anion-3}\label{dimer-anion-3}
\end{subfigure}

\caption{ Optimized B3LYP/6-31G(2df,p) structure for Dimer of Acrylic acid and its possible conjugate bases. Color code: Red-Oxygen, Grey-Hydrogen, Black-Carbon. }
\label{dimer}
\end{figure}

\begin{figure}[h]

\begin{subfigure}{.3\textwidth}
  \includegraphics[scale=.20,clip=true, trim = 10mm 10mm 10mm 30mm]{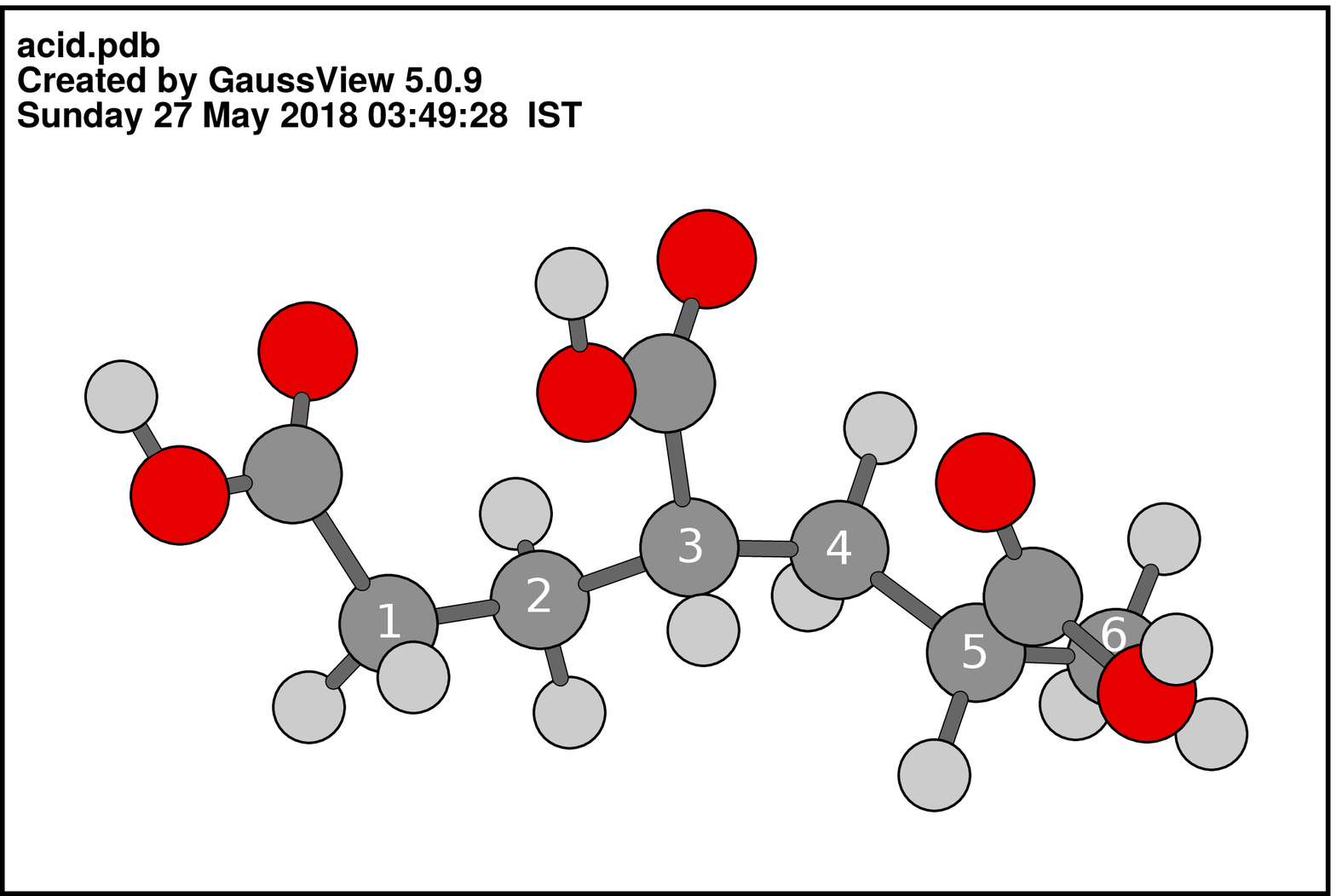} 
   \caption{Acid}\label{trimer_acid}
\end{subfigure}
\begin{subfigure}{.3\textwidth}  
  \includegraphics[scale=.20,clip=true, trim = 10mm 10mm 10mm 30mm]{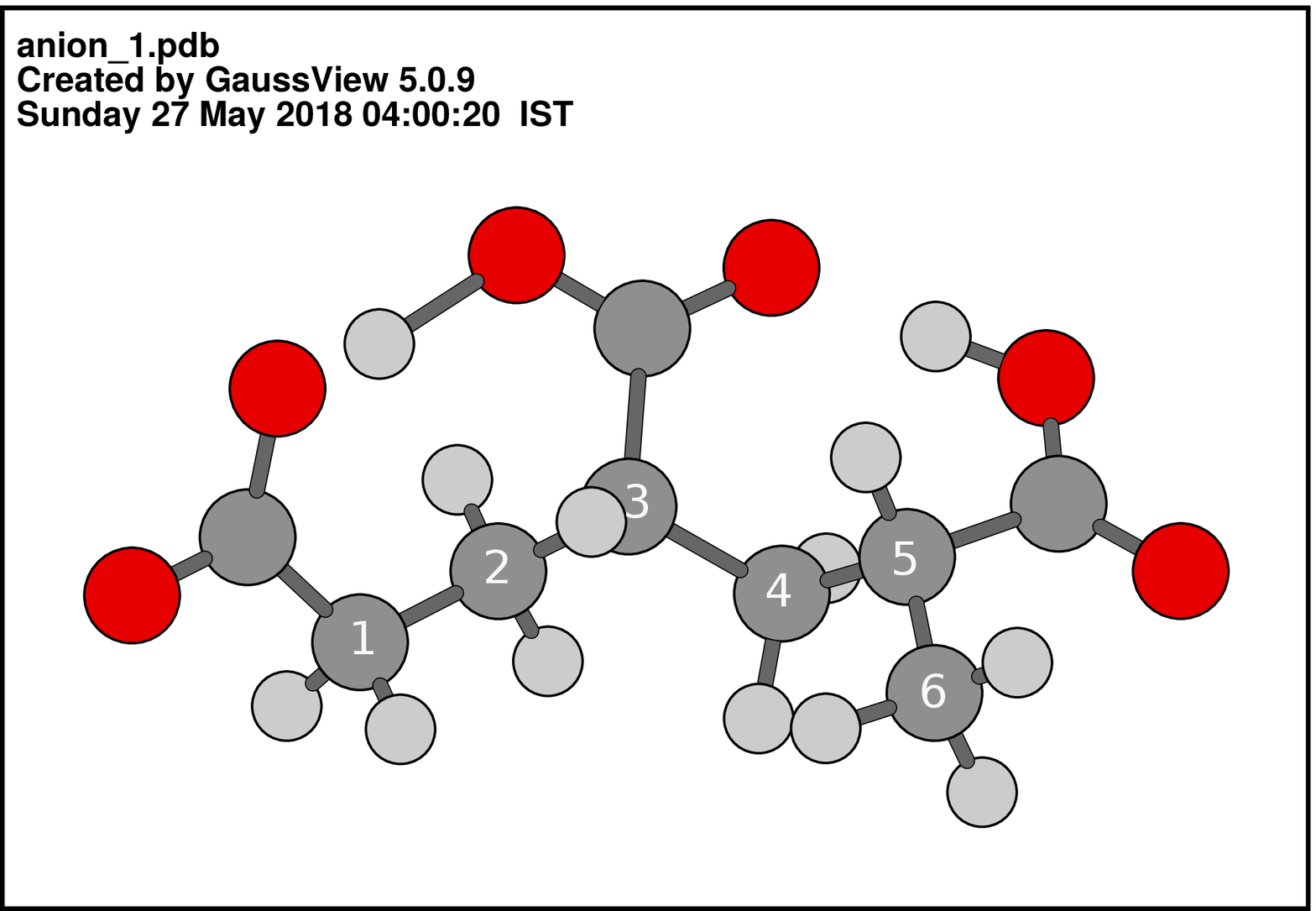}
   \caption{Anion-1}\label{trimer-anion-1}
\end{subfigure}
\begin{subfigure}{.3\textwidth}
\includegraphics[scale=.20,clip=true, trim = 10mm 10mm 10mm 30mm]{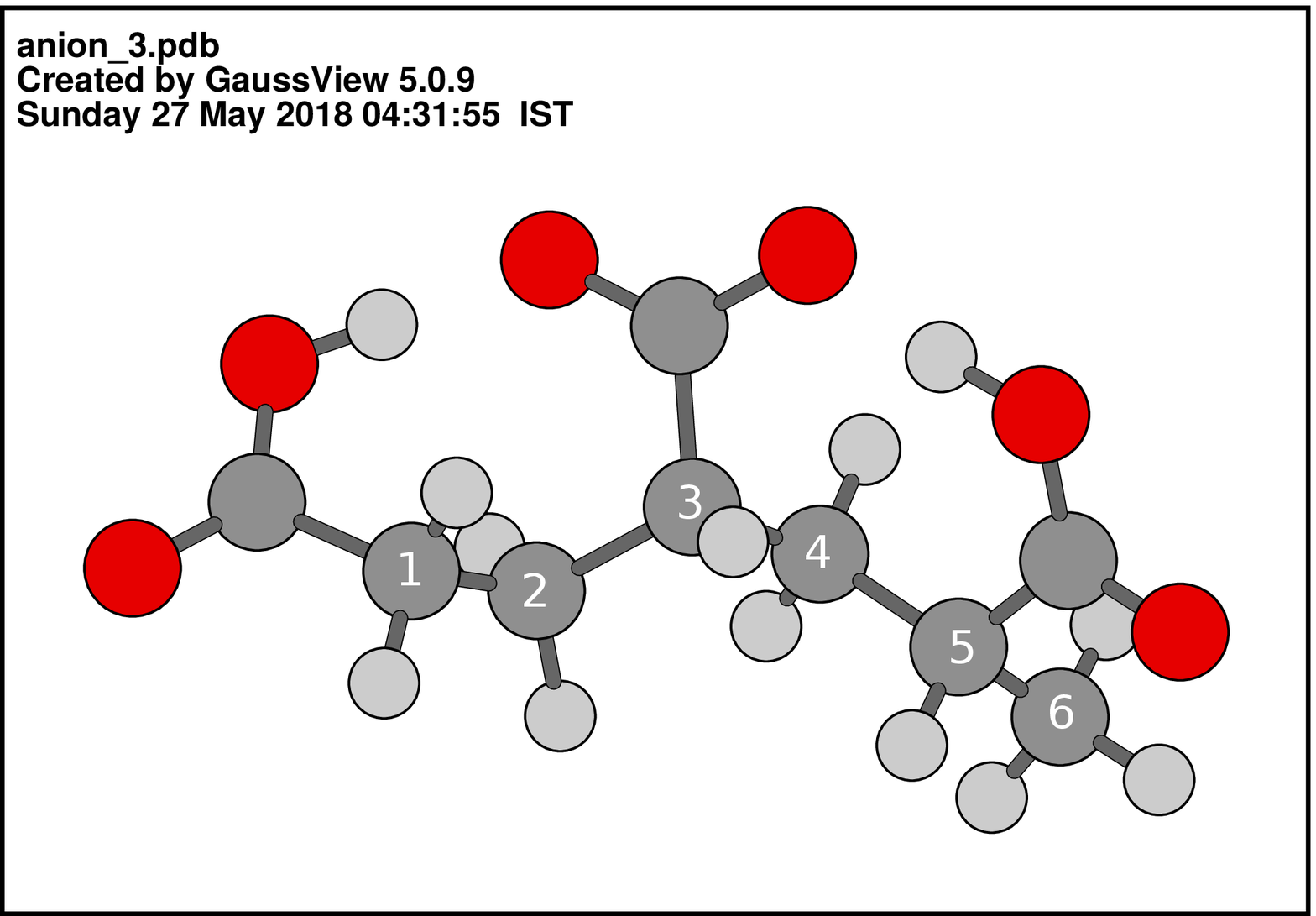}
 \caption{Anion-3}\label{trimer-anion-3}
\end{subfigure}
\begin{subfigure}{.3\textwidth}
\includegraphics[scale=.20,clip=true, trim = 10mm 10mm 10mm 30mm]{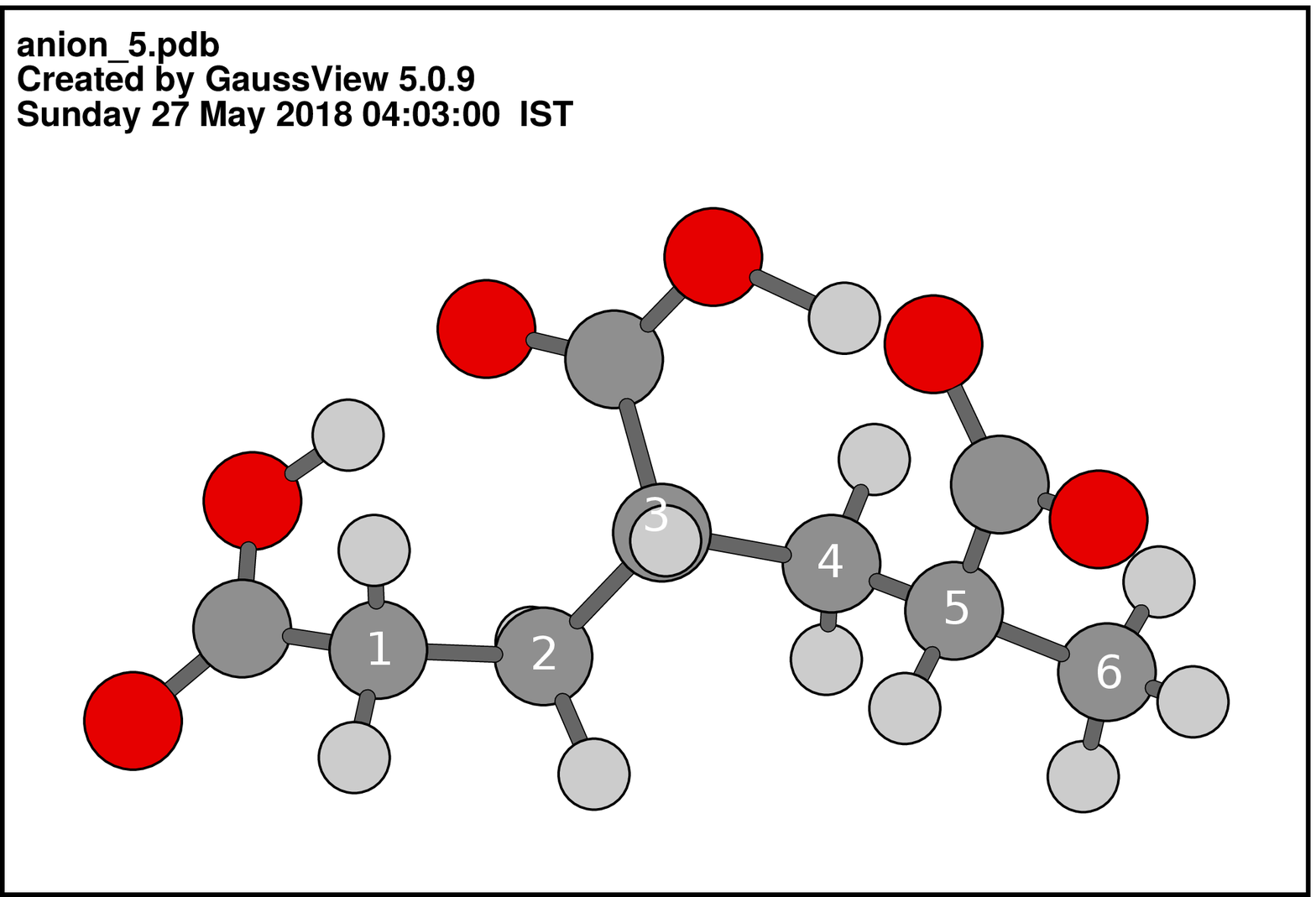}
 \caption{Anion-5}\label{trimer-anion-5}
\end{subfigure}

\caption{ Optimized B3LYP/6-31G(2df,p) structure for Trimer of Acrylic acid and its possible conjugate bases. Color code: Red-Oxygen, Grey-Hydrogen, Black-Carbon.}
\label{trimer}
\end{figure}

Table~\ref{pka} summarizes the gas-phase, solvation free energies and pka values for the oligomers of Acrylic acid. Figure~\ref{dimer} and figure~\ref{trimer} shows the optimized structures of dimer, trimer and their conjugate base structures obtained from CBS-QB3 gas phase calculations. Optimizing the geometry by other methods such as DFT/B3LYP(2df,p) and HF/6-31G(d) in gas phases and using SMD~\cite{SMD} and CPCM solvent models for solution phase respectively yielded similar stuctures. We optimized the geometries from different starting points. In certain optimized structutres, the intramolecular hydrogen-bonding was not observed, but since their energies were higher than the corresponding structures with H-bonds, they were not considered. 

The initial structures of conjugate bases of the dimer was obtained by deleting the Hydrogen atom at the 1, 3 positions. After optimizing these structures, we arrive at a quite similar structures of the anion-1 and anion-3. This shows that, in an experimental system, the deprotonated site cannot be determined because the molecule will optimize itself to the intramolecular H-bonded structure which is not dependent on which of the differentlly positioned COOH group was deprotonated.

It should be noted that a slight variation in the structure gives very different values of energy.

An intramolecular hydrogen-bonding can be seen clearly for all anionic structures, leading to the molecule stabilization.  

It is interesting to note that in figure~\ref{trimer-anion-3}, the deprotonated carboxylic group is accepting two hydrogen bonds from the both sides. This leads to increased stabilization of this anionic structure as compared to those in figure~\ref{trimer-anion-1} and ~\ref{trimer-anion-5}.

\begin{figure}[h]

\begin{subfigure}{.3\textwidth}
  \includegraphics[scale=.20,clip=true, trim = 10mm 10mm 10mm 30mm]{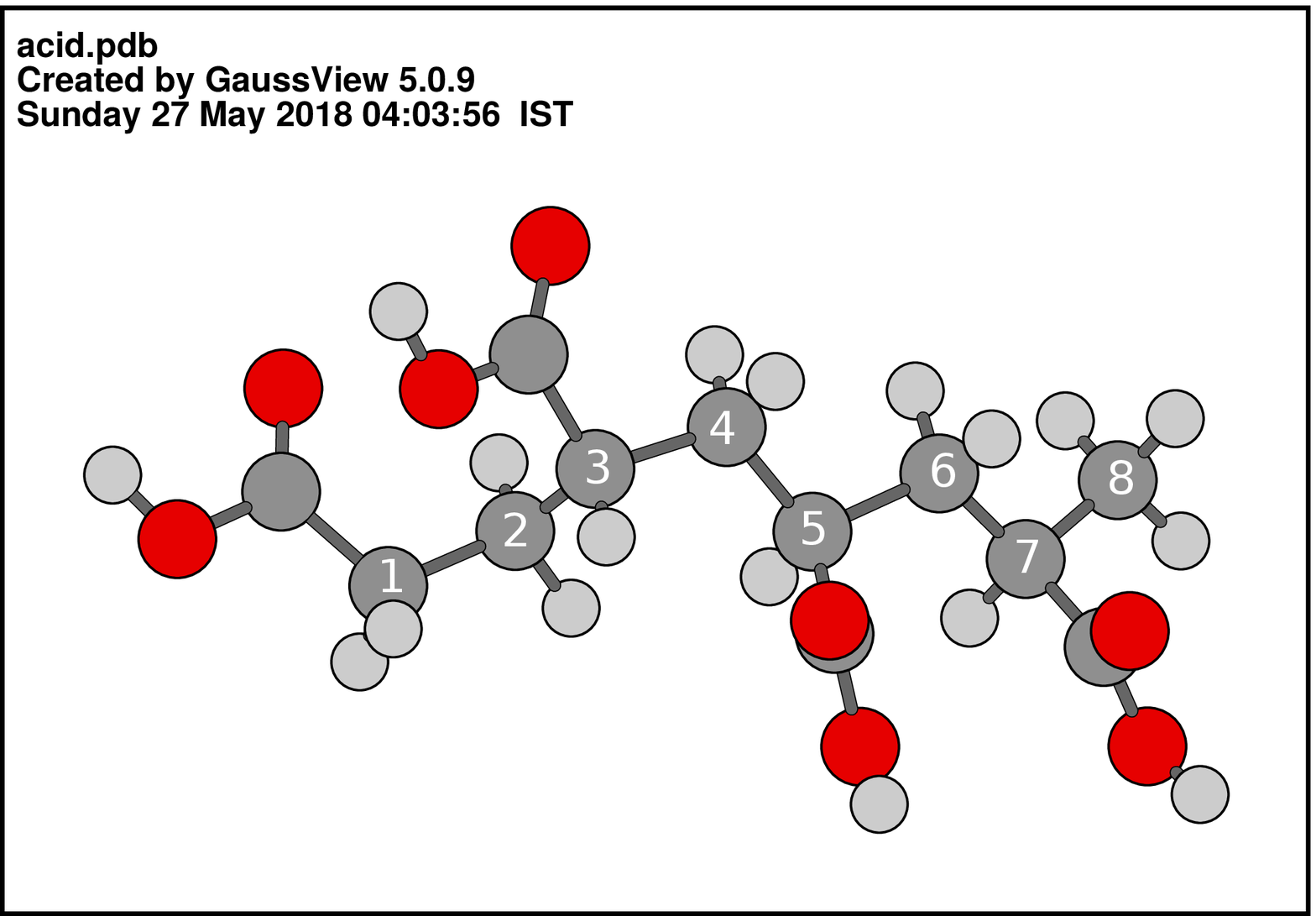} 
   \caption{Acid}\label{tetramer_acid}
\end{subfigure}
\begin{subfigure}{.3\textwidth}  
  \includegraphics[scale=.20,clip=true, trim = 10mm 10mm 10mm 30mm]{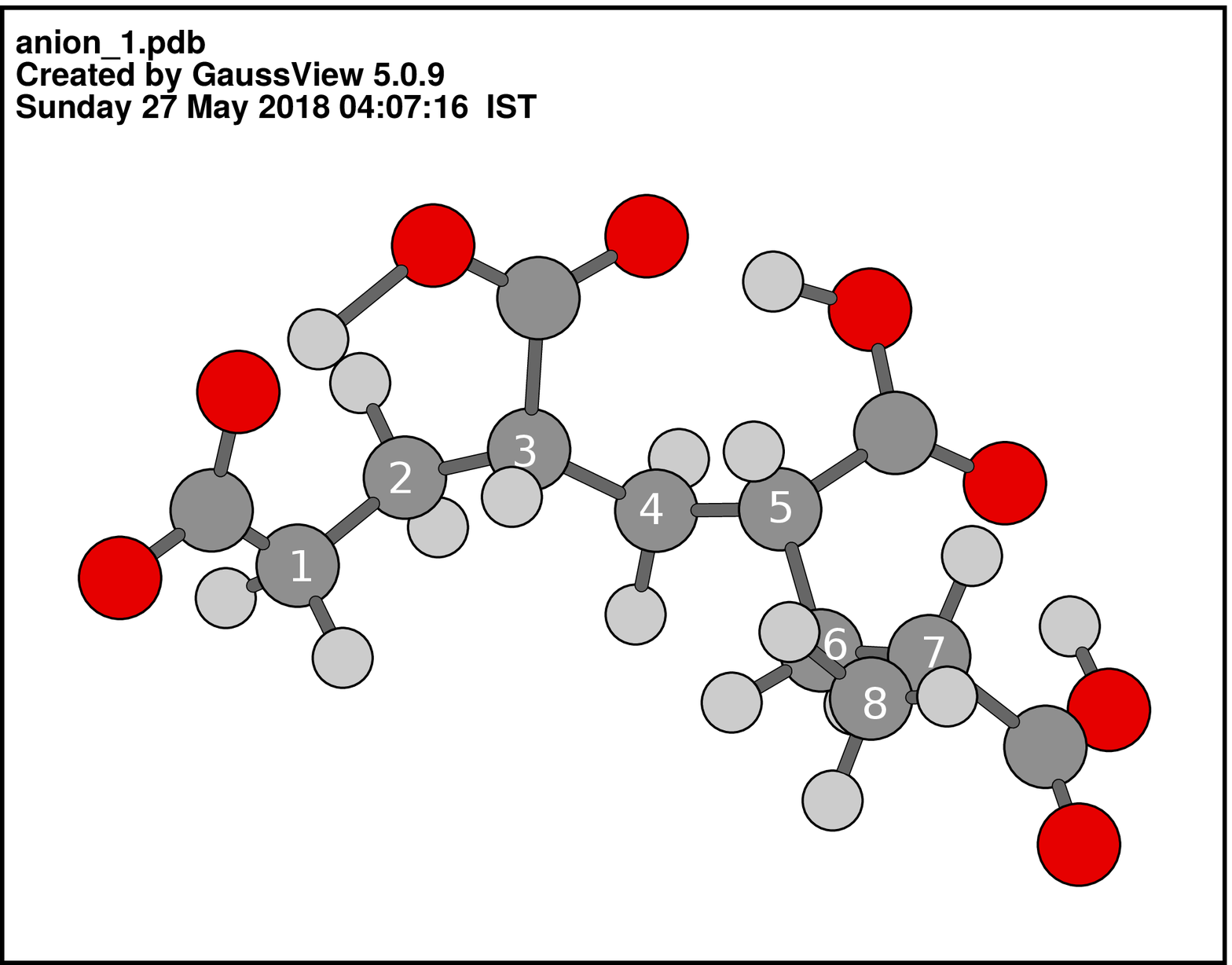}
   \caption{Anion-1}\label{tetramer-anion-1}
\end{subfigure}
\begin{subfigure}{.3\textwidth}
\includegraphics[scale=.20,clip=true, trim = 10mm 10mm 10mm 30mm]{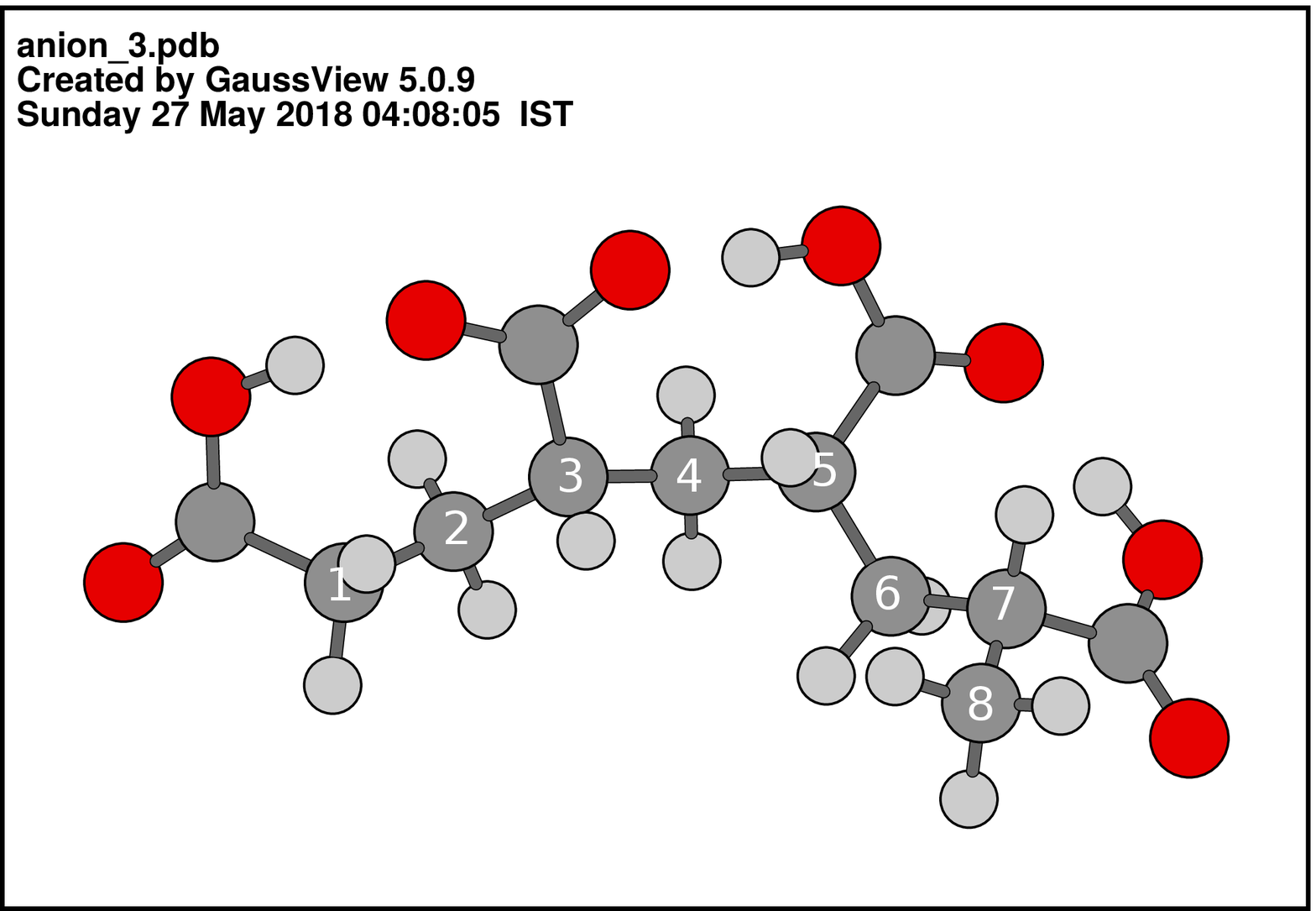}
 \caption{Anion-3}\label{tetramer-anion-3}
\end{subfigure}
\begin{subfigure}{.3\textwidth}
\includegraphics[scale=.20,clip=true, trim = 10mm 10mm 10mm 30mm]{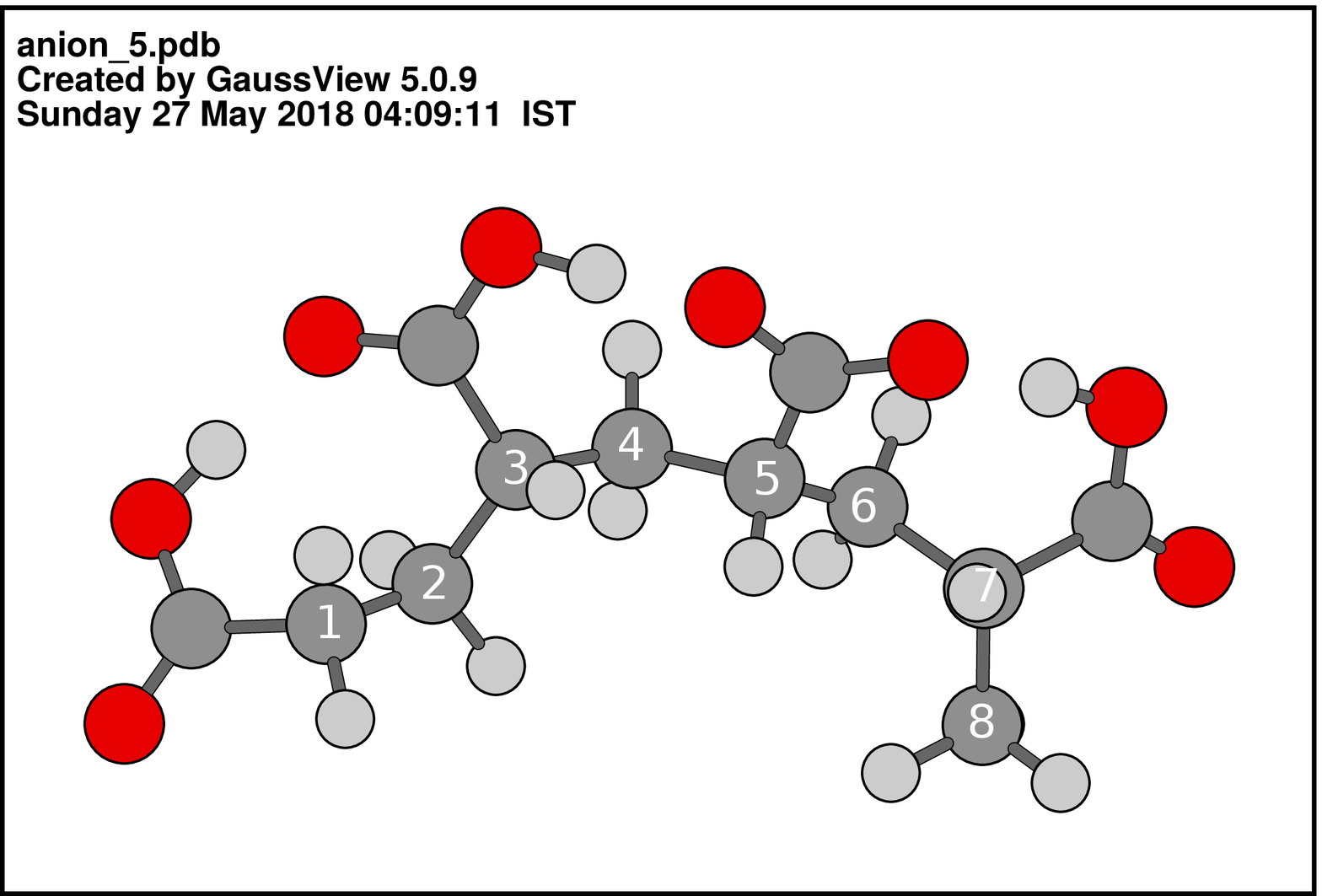}
 \caption{Anion-5}\label{tetramer-anion-5}
\end{subfigure}
\begin{subfigure}{.3\textwidth}
\includegraphics[scale=.20,clip=true, trim = 10mm 10mm 10mm 30mm]{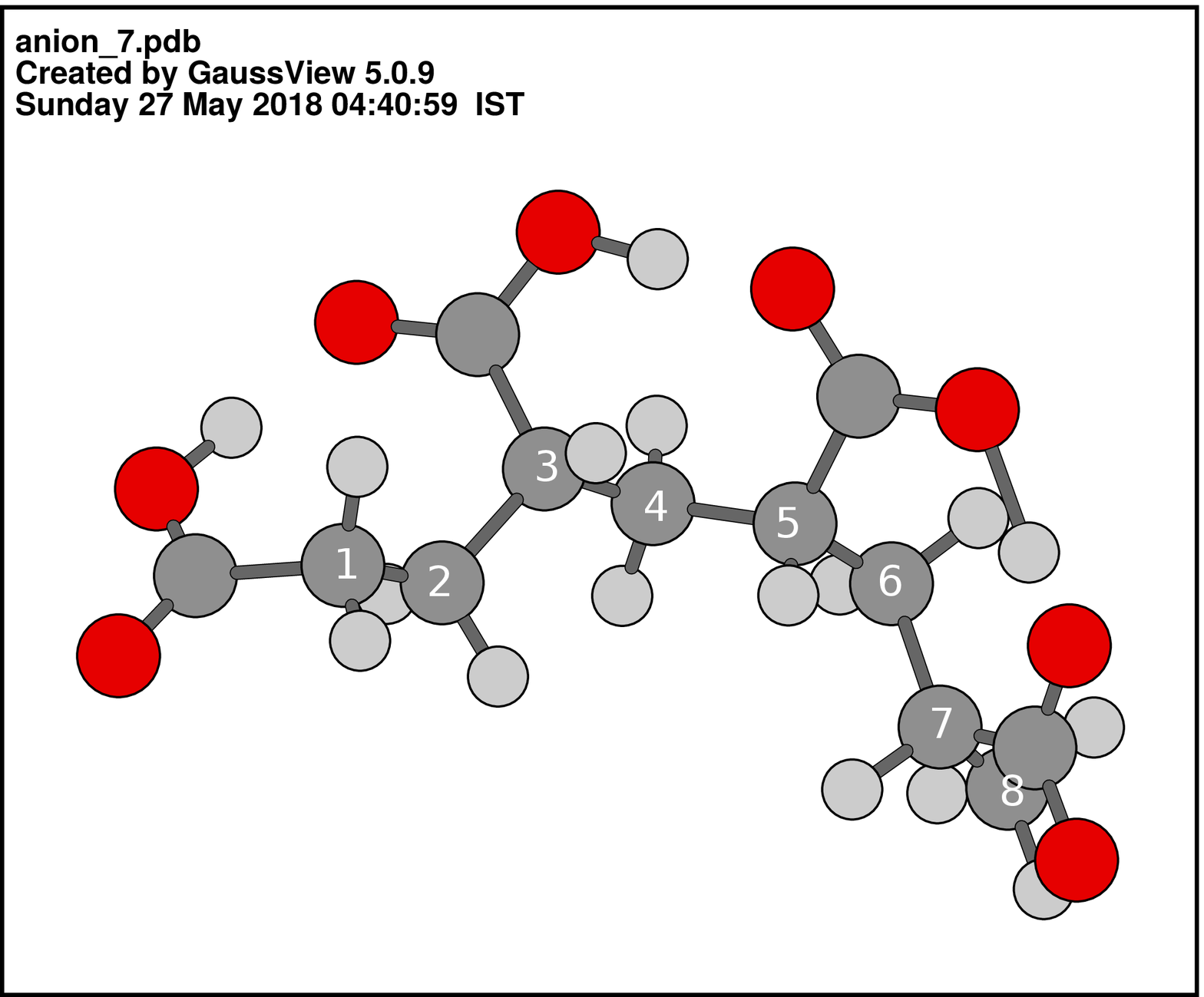}
 \caption{Anion-5}\label{tetramer-anion-7}
\end{subfigure}

\caption{ Optimized B3LYP/6-31G(2df,p) structure for Tetramer of Acrylic acid and its possible conjugate bases. Color code: Red-Oxygen, Grey-Hydrogen, Black-Carbon.}
\label{tetramer}
\end{figure}

In all the anionic structures in figures~\ref{dimer},~\ref{trimer} and ~\ref{tetramer}, it was observed that all the carboxylic acids are participating in intramolecular H-bonding within the molecule. This leads to the formation of a helix-type structure if all the O are joined by an imaginary line. The same thing is not true for acid in which intramoleculer H-bonding was not observed. It is also interesting to observe that the orientation of OH group in COOH is changed to anti when the group donates the Hydrogen to the COO$^-$ group. This is because such orientation provides easier formation of H-bonds.

\begin{figure}[h!]

  \includegraphics[scale=.20,clip=true, trim = 10mm 10mm 10mm 30mm]{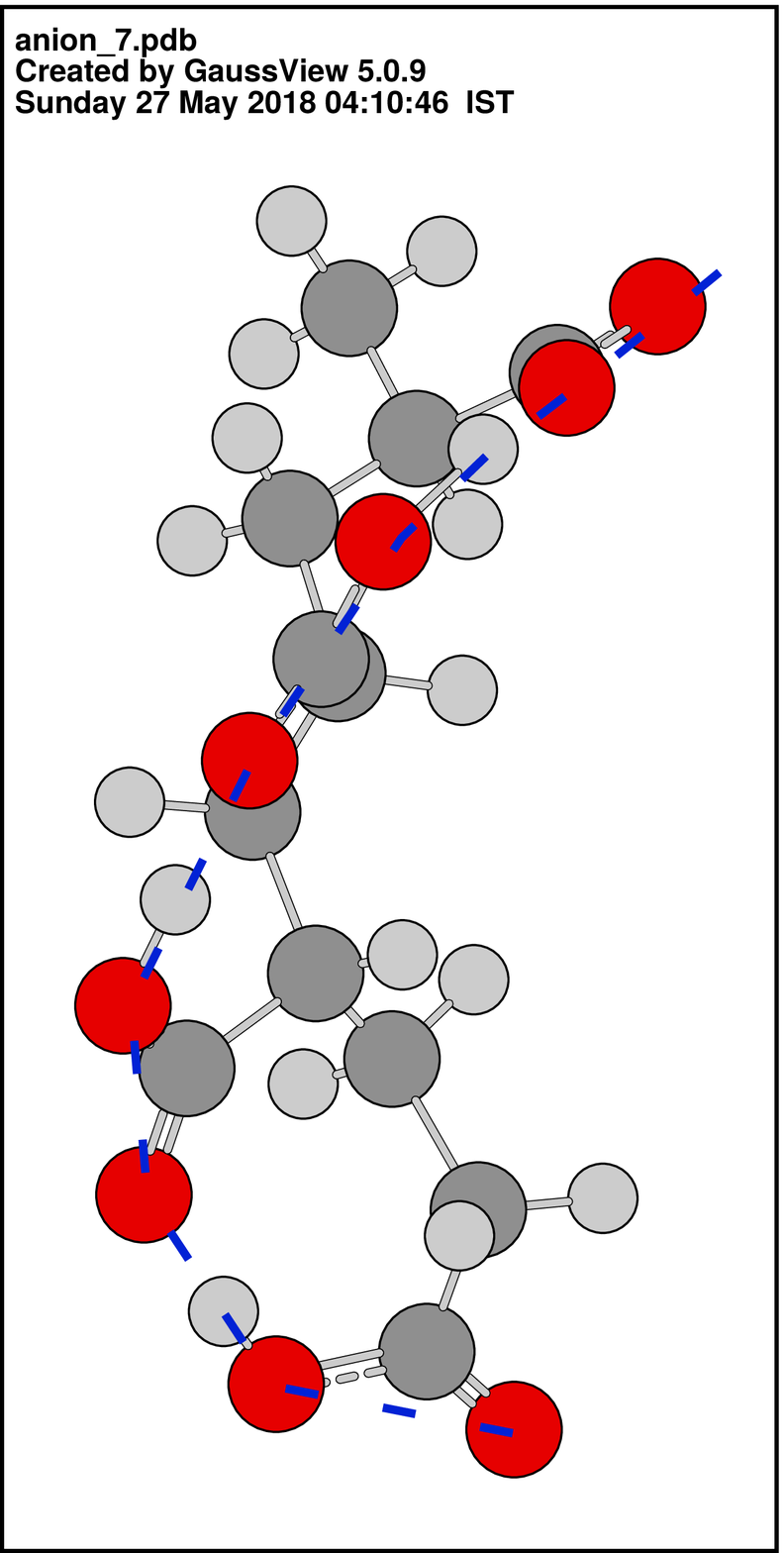} 
   \caption{Optimized B3LYP/6-31G(2df,p) structure for anion-7. All the O atoms are on the path of a helix as shown by a dotted line. }
\label{helix}
\end{figure}

In polymer physics, the Radius of gyration is used to approximate the dimension of a polymer chain. It is defined as the avergae distance of any atom from the center of mass of the polymer chain. The square of radius of gyration is defined as - 
\begin{equation}
R_g^2 = \frac{1}{N} \sum^N_{k=1}(\vec{r}_k-\vec{r}_{cm})^2
\label{rg}
\end{equation}

Where N is the number of atoms in the chain, $\vec{r}_k$ is the position vector of the $k^{th}$ atom, and $\vec{r}_{cm}$ is the position vector of the center of mass of the chain which is given by -

\begin{equation}
\vec{r}_{cm} = \frac{1}{N} \sum^N_{k=1}\vec{r}_k
\end{equation}
Radius of gyration has been computed by considering only the carbon atoms that forms the structure and neglecting the carbon from COOH groups. The values calculated using equation~\ref{rg} has been reported in table~\ref{Rg}. It can be concluded from the table~\ref{Rg} that $R_g$ for each molecule is slightly less in presence of solvent field as compared to its gaseous optimized structure. This might not be true in case we take explicit solvents. This is because using implicit solvent fields, the H-bonding that may exist water and COO$^-$ group which is not captured fully here.

\begin{table}[h!]
\centering
\begin{tabular}{c| c |c | c |c |c}

 \hline
 \multirow{2}{4em}{Molecule}
 & \multicolumn{2}{|c|}{Single deprotonation } &  \multirow{2}{6em}{Molecule} & \multicolumn{2}{|c|}{Second deprotonation} \\
 
 & Gas & SMD  &  & Gas   & SMD  \\
 \hline
\textbf{Dimer}(acid) & 1.4974 & 1.4958 & \textbf{Dimer}(Anion-1\&3) & 1.91075 & 1.91078  \\
\textbf{Dimer}(Anion-1) & 1.4971 & 1.4497 & \textbf{Trimer}(Anion-1\&3) & 2.30965 & 2.27497 \\
\textbf{Dimer}(Anion-3) & 1.4972 & 1.8223 & \textbf{Trimer}(Anion-3\&5) & 2.43704 & 2.39587   \\ 
\textbf{Trimer}(Acid) & 2.1851 & 2.2074 & \textbf{Trimer}(Anion-5\&1) & 2.37316 & 2.30754 \\
\textbf{Trimer}(Anion-1) & 2.0863 & 2.0809 & \textbf{Tetramer}(Anion-1\&3) & 2.47130 & 2.52291 \\
\textbf{Trimer}(Anion-3) & 2.1614 & 2.1562 & \textbf{Tetramer}(Anion-1\&5) & 2.59831 & 2.55671 \\
\textbf{Trimer}(Anion-5) & 2.1615 & 2.1562 & \textbf{Tetramer}(Anion-1\&7) & 2.60598 & 2.56137 \\
\textbf{Tetramer}(Acid) & - & - & \textbf{Tetramer}(Anion-3\&5) & 2.71293 & 2.72355 \\
\textbf{Tetramer}(Anion-1) & 2.6647 & 2.6533 & \textbf{Tetramer}(Anion-3\&7) & 2.72706 & 2.72653 \\
\textbf{Tetramer}(Anion-3) & 2.6903 & 2.6763 & \textbf{Tetramer}(Anion-5\&7) & 2.55338 & 2.53800 \\
\textbf{Tetramer}(Anion-5) & 2.6890 & 2.6689 & - & - & - \\
\textbf{Tetramer}(Anion-7) & 2.7428 & 2.7279 & - & - & - \\

 \hline
 
\end{tabular}
\caption{Radius of gyration (in \AA) for each molecule was calculated using the Carbon atom of the structural frame.}
\label{Rg}
\end{table}

\section{Conclusion}

Our theoretical calculations lack support from the experimental data. However, no direct experimental measurements are possible for the solvation free energies for  a single ion.~\cite{Dupont}
second source of uncertainty comes from the lack of
experimental data for ions.

Although though our simulation, the optimized structures and therefore the energies of the conjugate bases for a particular molecule is different(depending on the position of COOH undergoing the deprotonation), in reality it might be true that the final optimized structure might not depend on the position. There is a possibility that all these structres might be interconvertible, each having a different probablity of existing depending on its electronic energy values. This picture unfortunately could not be captured via the so-performed quantum mechanical optimization because each theory we have used in this study has its own limitations, and most of them are generally do not give best results when applied to larger molecules that were considered here.

Our study opens the scope for improvement of quantum mechanical methods to become computationally efficient and accurate for large molecules. This study is just a step forward in predicting the acidic properties of molecules with multiple acidic sites and their behavior at different degree of deprotonation.

\clearpage

\bibliography{ref}

\end{document}